\documentclass[showpacs,preprintnumbers,amsmath,amssymb,prl,twocolumn]{revtex4}

\usepackage{graphicx}
\usepackage{dcolumn}
\usepackage{bm}
\input epsf.tex

\def\comment#1{}

\def\iems#1{}
\def\ins#1{}

\begin{document}

\title{Boltzmann Distribution and Market Temperature
}
\author{H. Kleinert}
\email{kleinert@physik.fu-berlin.de}
\author{X.J. Chen}
\email{xiaojiang.tim@googlemail.com}
\affiliation{Institut f{\"u}r Theoretische Physik,
Freie Universit{\"a}t Berlin, Arnimallee 14, D-14195 Berlin, Germany\\
Nomura House, 1 St Martin's-le-Grand,
London EC1A 4NP, UK
}

\begin{abstract}
The minute fluctuations of of S\&P 500 and NASDAQ 100 indices
display Boltzmann statistics  over a wide range
of positive as well as negative returns, thus allowing us to define
a {\em market temperature\/} for either sign. With
increasing time the sharp
Boltzmann peak broadens into a Gaussian whose volatility $ \sigma $
measured in $1/ \sqrt{ {\rm min}}$ is related to the temperature $T$ by $T=
\sigma / \sqrt{2}$. Plots over the years 1990--2006 show that the
arrival of the 2000 crash was preceded by an increase in
market temperature, suggesting that this increase can be used
as a warning signal for crashes.
A  plot of the Dow Jones temperature
over 78 years reveals a remarkable stability through many historical turmoils,
interrupted only by short heat bursts near the crashes.
\end{abstract}

\maketitle

It is by now well-known that financial data do no display Gaussian
distributions \cite{MS1,BC1,Pott1,Pott2,Sornette,Lux96,SorMal,Y1,TSALLIS,TSALLISN,PJ,Voigt,PI}. Most importantly, the tails of the
distributions are power-like \cite{Go+98}, since large fluctuations are much more
frequent than in a Gaussian distribution. This is of great
importance for financial institutions who want to  estimate the risk
of market crashes.

In this note we would like to focus on the opposite regime of the
most frequent events near the peak of the distribution.
The logarithms of the
stock prices $x(t)=\log S(t)$ and thus also
 of NASDAQ 100 and S\&P 500 indices
have a special property: the minute returns
 $z(t)=  \Delta  x(t)$
show an exponential distribution \cite{EXPOD}
for positive as well as negative $z(t)$, as long as the probability is
rather large \cite{Y2,Y4}.

\begin{equation}
\begin{array}{ccc}
\tilde B(z)=\displaystyle\frac{1}{2T}e^{-|z|/T}.
\end{array}
\label{doubleexponential}
\end{equation}
\def\IncludeEpsImg#1#2#3#4{\renewcommand{\epsfsize}[2]{#3##1}{\epsfbox{#4}}}
\begin{figure}[h]
\unitlength.5mm
\begin{tabular}{lll}
\begin{picture}(0,76)
\put(-90,-10){\IncludeEpsImg{19.02mm}{16.09mm}{0.4500}{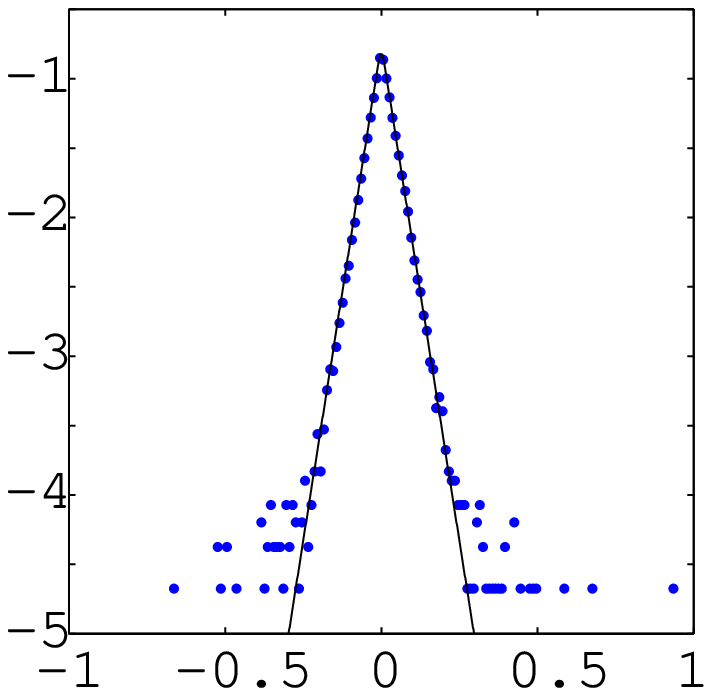}}
\put(10,-10){\IncludeEpsImg{19.02mm}{16.09mm}{0.4500}{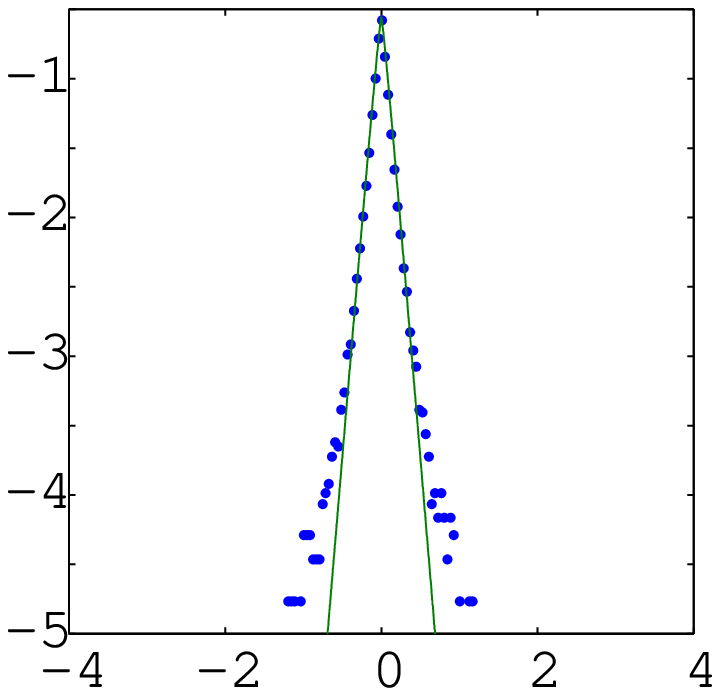}}
\put(-77,68){\scriptsize\rm{S\&P500 2004-05(1min)}}
\put(24,68){\hspace{-1.1em}\scriptsize\rm{NASDAQ100 2001-02(1min)}}
\put(-75,57){\scriptsize$\log P(z)$} \put(25,57)
{\scriptsize$\log P(z)$} \put(-63.5,-1){\scriptsize$z$ in percent}
\put(36.5,-1){\scriptsize$z$ in percent}
\end{picture}
\end{tabular}
\caption{Boltzmann distribution
of minute returns of S\&P 500
and NASDAQ 100 indices.} \label{1min}
\end{figure}

In Fig.~\ref{1min}
we show that the data are fitted well by the
distribution \cite{Yahoo}. Only a very small set of rare
events of large $|z|$ does not follow the exponential law, but
displays heavy tails.
If the exponential distribution
is interpreted as a Boltzmann
distribution,
the parameter $T$ in
(\ref{doubleexponential})
plays the role of a
market temperature, and there are statistical
considerations to support this interpretation
 \cite{Aoki,Y3}.
The purpose of this note
is to determine the market temperatures
for the
S\&P 500
and NASDAQ indices over
many years.

In principle, there are different  temperature $T_{\pm}$ for positive and
negative returns, but to a good approximations
we may equate both $T\approx T_+\approx T_-$.

\begin{figure}[h]
\begin{tabular}{lll}
\begin{picture}(10,95)
\put(-120,-7){\IncludeEpsImg{19.02mm}{16.09mm}{0.4500}{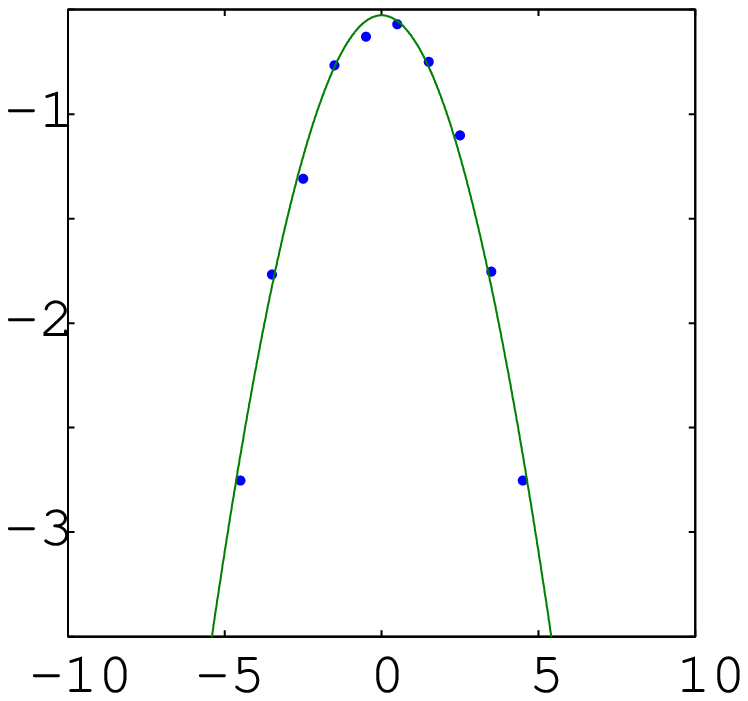}}
\put(10,-7){\IncludeEpsImg{19.02mm}{16.09mm}{0.4500}{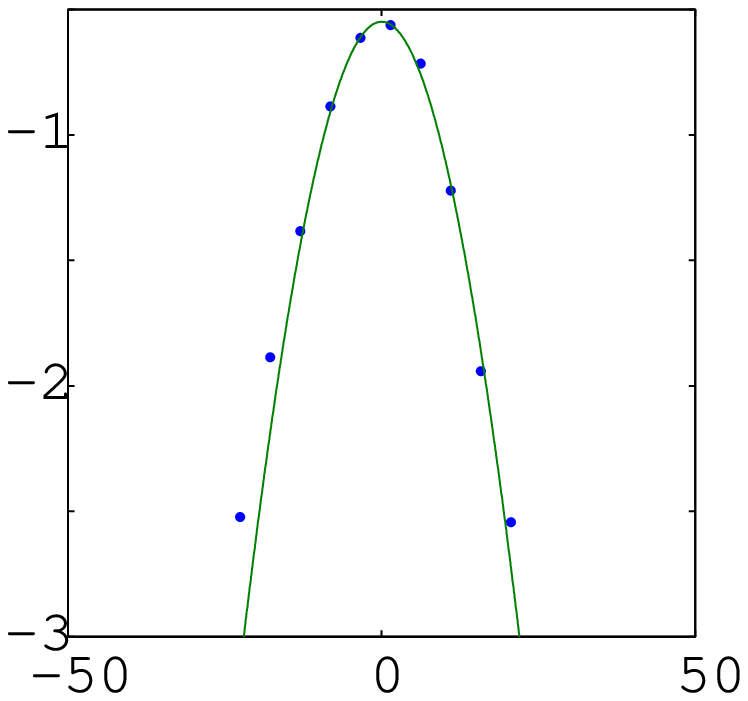}}
\put(-103,92){\scriptsize{\rm{S\&P500 2004-05(1week)} }}
\put(29,92){\hspace{-1.1em}\scriptsize\rm{NASDAQ100 2001-02(1week)} }
\put(-100,77){\scriptsize$\log P(z)$} \put(30,77){\scriptsize$\log
P(z)$} \put(-81,-10){\scriptsize$z$ in percent}
\put(49,-10){\scriptsize$z$ in percent}
\end{picture}
\end{tabular}
\caption{Gaussian distributions of S\&P 500 and
NASDAQ 100 weekly returns.} \label{gaussdata}
\end{figure}

At larger time scales, the distribution becomes more and more
Gaussian, as required by the {\em central limiting theorem\/}
\iems{central+limiting+theorem}%
\iems{theorem,central+limiting}%
\iems{limiting+theorem;central}%
of statistical mechanics which states that the convolution of
infinitely many arbitrary distribution functions of finite width
always approaches  a  Gaussian distribution. This is illustrated by
the the weekly data of the two indexes in Fig. \ref{gaussdata}.

\begin{figure}
\begin{tabular}{lll}
\begin{picture}(10,88)
\put(-126,-5){\IncludeEpsImg{19.02mm}{16.09mm}{0.3500}{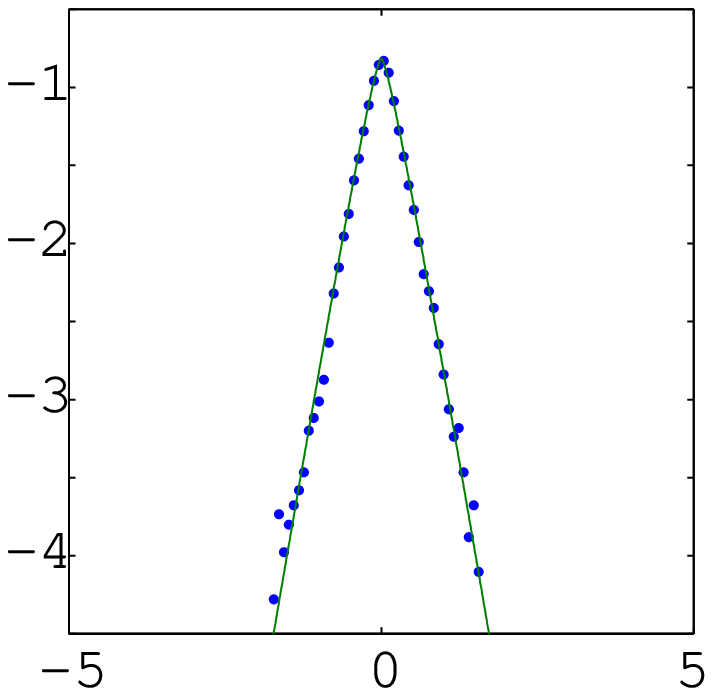}}
\put(-40,-5){\IncludeEpsImg{19.02mm}{16.09mm}{0.3500}{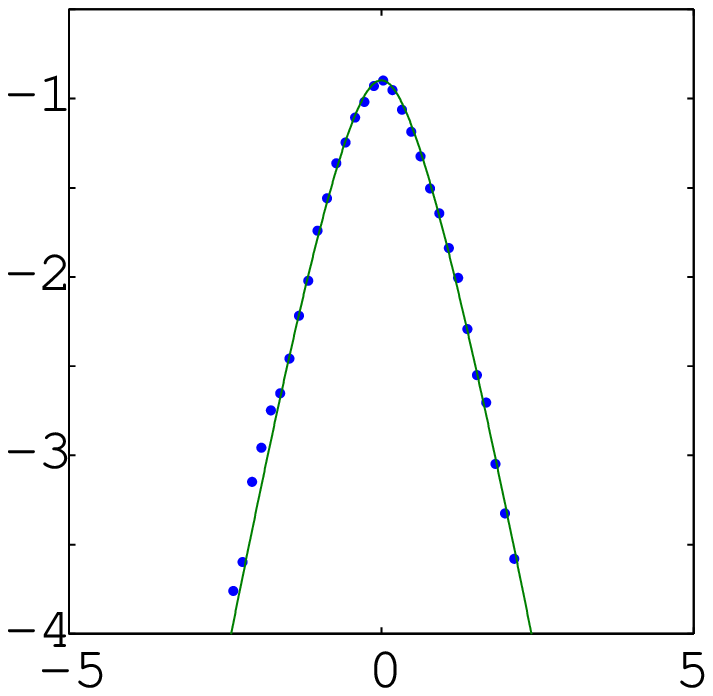}}
\put(46,-5){\IncludeEpsImg{19.02mm}{16.09mm}{0.3500}{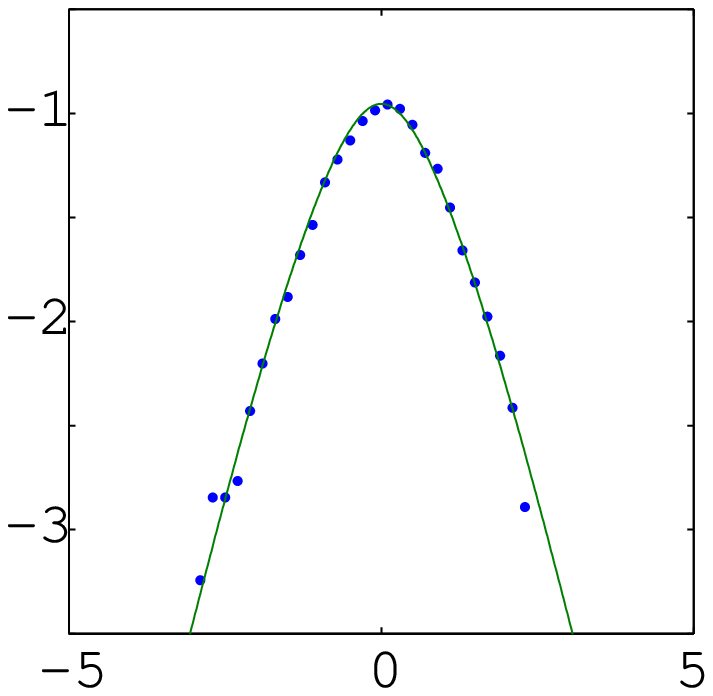}}
\put(-111,70){\scriptsize$\log P(z)$} \put(-25,70){\scriptsize$\log
P(z)$} \put(61,70){\scriptsize$\log P(z)$}
\put(-101,4){\scriptsize$z$ in percent} \put(-14,4){\scriptsize$x$
in percent} \put(71,4){\scriptsize$z$ in percent}
\put(-86,-7){\scriptsize\rm{(a)}}\put(1,-7){\scriptsize\rm{(b)}}\put(87,-7){\scriptsize\rm{(c)}}
\end{picture}
\end{tabular}
\caption{Fits of convolution of Boltzmann distribution to  S\&P 500
returns in Fig.~1 over time intervals of 1 hour,  4 hours, and 1 day,
respectively.} \label{transit}
\end{figure}

The transition from Boltzmann
to Gaussian distributions
is shown for the  S\&P 500 index in Fig.~\ref{transit}.

The convergence to a Gaussian distribution
is in contrast to the pure L\'evy distribution
of infinite width
where
the falloff
remains
power-like
at large distances for any data frequency.
This will happen here as well for the rare
events outside of the Boltzmann regime.

\begin{figure}[b]
 \begin{picture}(1,266)
\put(-120,165){\IncludeEpsImg{194.02mm}{163.09mm}{0.450}{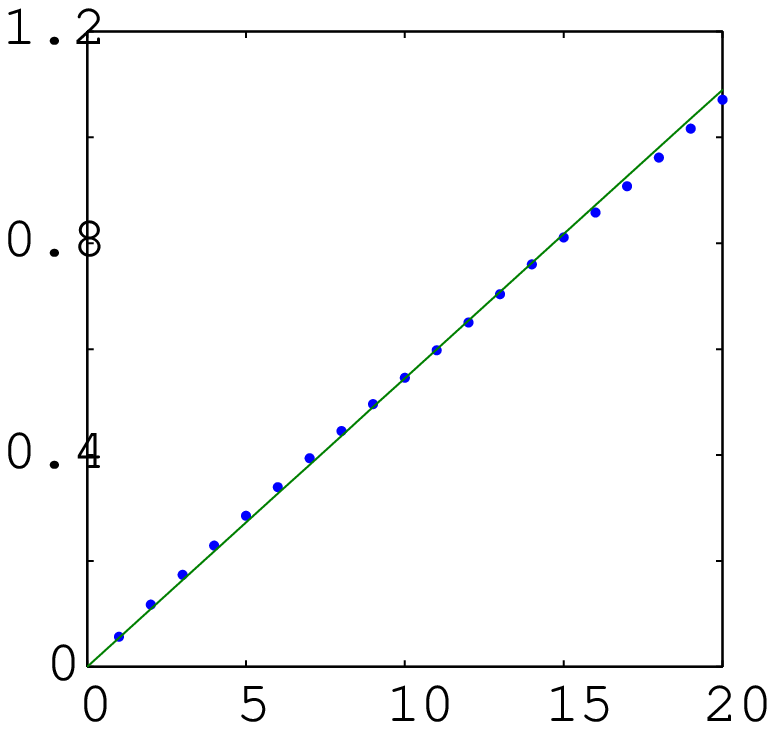}}
\put(0,165){\IncludeEpsImg{194.02mm}{163.09mm}{0.450}{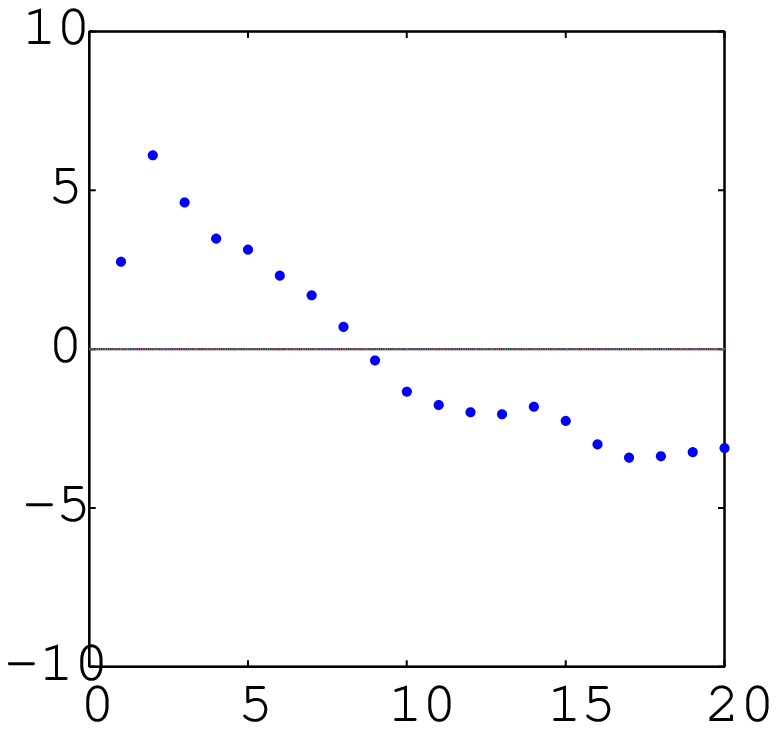}}
\put(-120,40){\IncludeEpsImg{194.02mm}{163.09mm}{0.450}{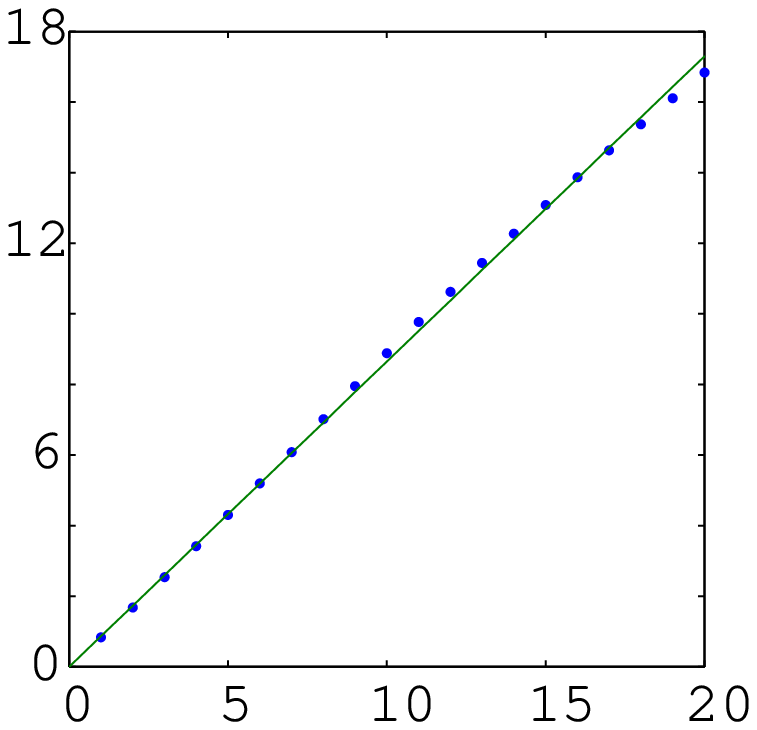}}
\put(0,40){\IncludeEpsImg{194.02mm}{163.09mm}{0.450}{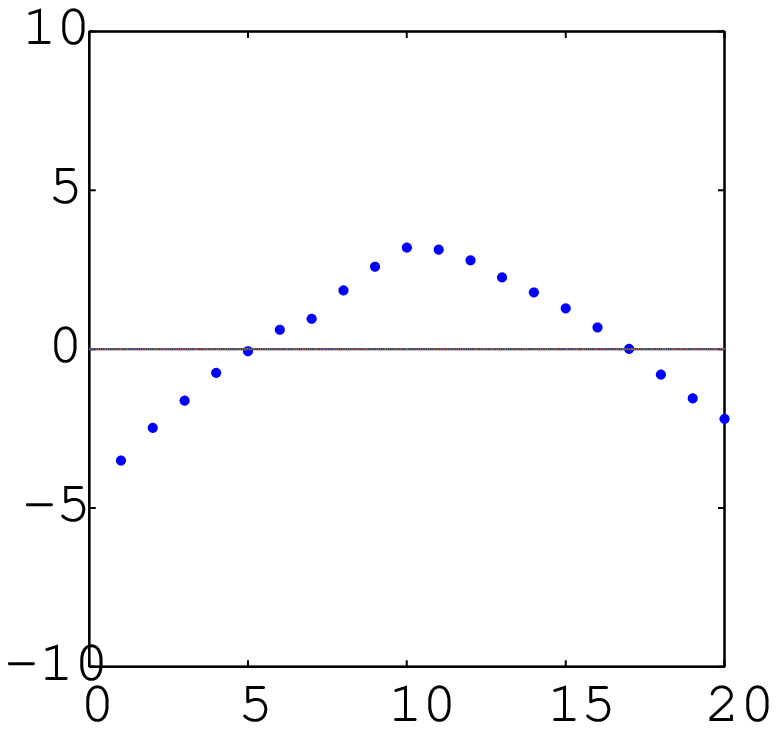}}
\end{picture}
\put(-93,265){\scriptsize\rm{S\&P500 2004-2005}}
\put(-86,140){\hspace{-1.1em}\scriptsize\rm{NASDAQ100 2001-02}}
\put(-95,247){\scriptsize$  \sigma  ^2(t)$}
\put(-95,120){\scriptsize$ \sigma ^2(t)$}
\put(-90,160){\scriptsize\rm{Time lag t (hours)}}
\put(-90,35){\scriptsize\rm{Time lag t (hours)}}
\put(28,265){\scriptsize\rm{S\&P500 2004-2005}}
\put(34,140){\hspace{-1.1em}\scriptsize\rm{NASDAQ100 2001-02}}
\put(20,247){\scriptsize\rm{~~~deviation in percent}}
\put(20,120){\scriptsize\rm{~~~deviation in percent}}
\put(28,160){\scriptsize\rm{Time lag t (hours)}}
\put(28,35){\scriptsize\rm{Time lag t (hours)}}
~\\[-2em]
 \caption{Variance of S\&P 500 and
 NASDAQ 100 indices
 as a function of time. The
right-hand side amplifies  the small relative deviation from the
linear shape in percent. } \label{spvariance}
\end{figure}%

\begin{figure}[tb]
 \begin{picture}(1,266)
\put(-120,165){\IncludeEpsImg{194.02mm}{163.09mm}{0.450}{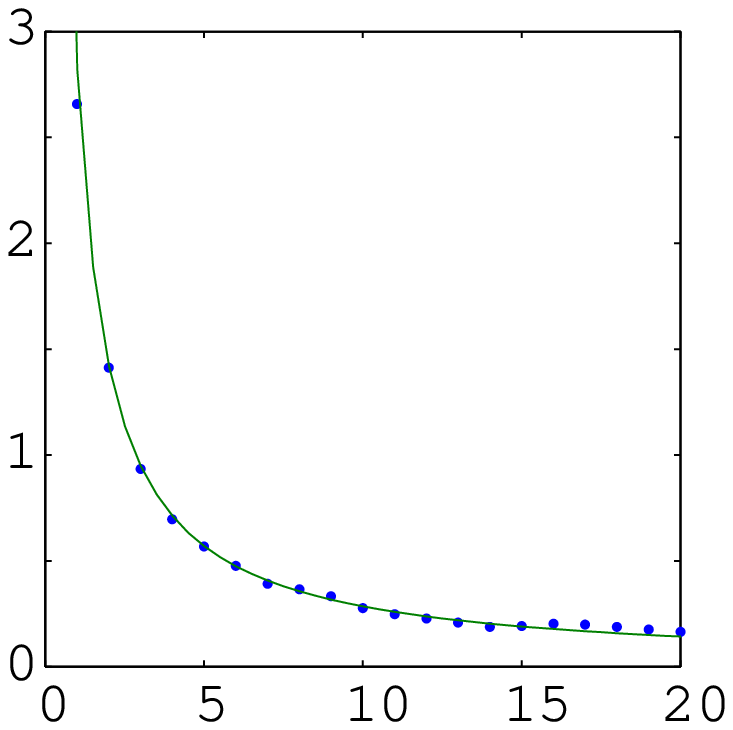}}
\put(0,165){\IncludeEpsImg{194.02mm}{163.09mm}{0.450}{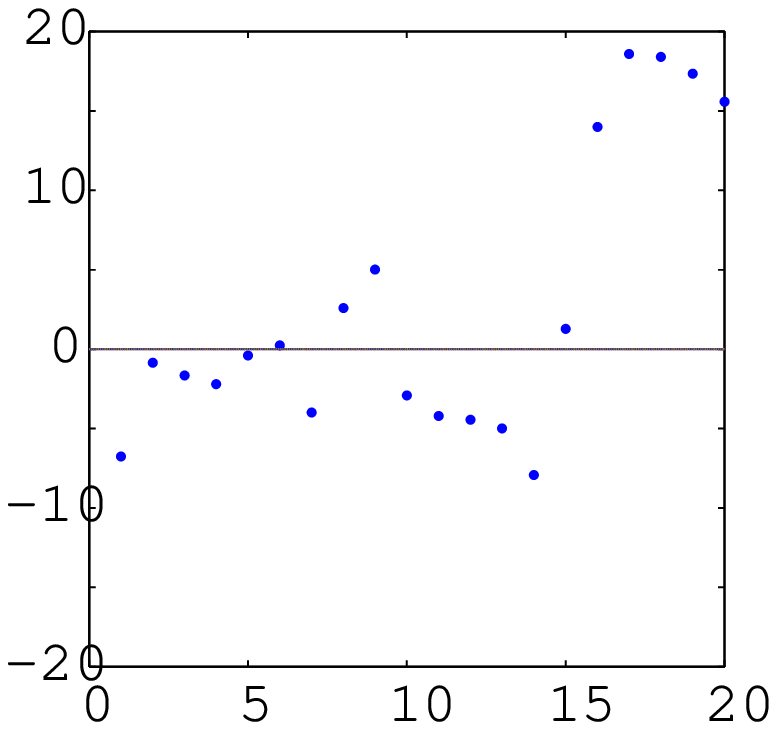}}
\put(-120,40){\IncludeEpsImg{194.02mm}{163.09mm}{0.450}{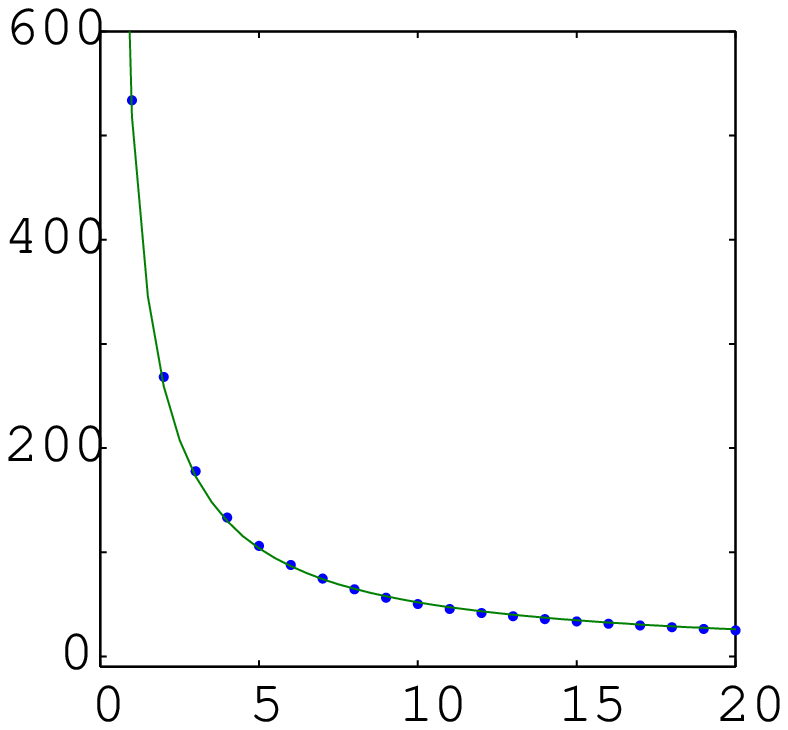}}
\put(0,40){\IncludeEpsImg{194.02mm}{163.09mm}{0.450}{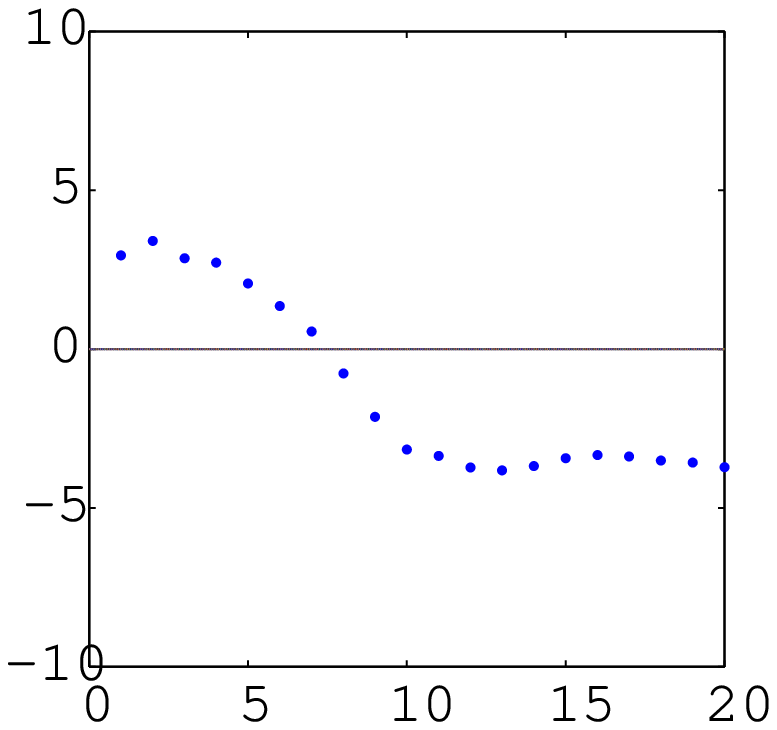}}
\end{picture}
\put(-93,265){\scriptsize\rm{S\&P500 2004-2005}}
\put(-86,140){\hspace{-1.1em}\scriptsize\rm{NASDAQ100 2001-02}}
\put(-95,237){\scriptsize$~~ \kappa (t)$}
\put(-95,110){\scriptsize$ ~~\kappa (t)$}
\put(-90,160){\scriptsize\rm{Time lag t (hours)}}
\put(-90,35){\scriptsize\rm{Time lag t (hours)}}
\put(28,265){\scriptsize\rm{S\&P500 2004-2005}}
\put(34,140){\hspace{-1.1em}\scriptsize\rm{NASDAQ100 2001-02}}
\put(20,247){\scriptsize\rm{~~~deviation in percent}}
\put(20,120){\scriptsize\rm{~~~deviation in percent}}
\put(28,160){\scriptsize\rm{Time lag t (hours)}}
\put(28,35){\scriptsize\rm{Time lag t (hours)}}
~\\[-2em]
 \caption{Kurtosis of S\&P 500 and
 NASDAQ 100 indices
 as a function of time. The
right-hand side shows the relative deviation from the $1/t$ behavior
in percent. } \label{spkappa}
\end{figure}%

The time dependence of the distribution is found in the usual way \cite{Pott2,PI}.
We calculate the Fourier transform of $B(z)$:
\\[-1.2em]

\begin{equation}
 B(p)=
\int_{-\infty}^\infty dx\, e^{ipz}\frac{1}{2T}e^{-|z|/T}=
\frac{1}{1+(Tp)^2}, \label{@}\end{equation}
and identify the Hamiltonian as
\begin{equation}
H(p)=\log[ 1+(Tp)^2].
\label{@}\end{equation}
This has only even  cumulants    $(
n=2,4,\dots)$:
\begin{equation}
c_n=-i^nH^{(n)}(0)=2i^n(-1)^{n/2}T^{n}(n-1)!.
\label{@ConnC}\end{equation}

As a function of time, the distribution widens as follows:
\begin{eqnarray}
\tilde B(z;t)&=&\int_{-\infty}^\infty
 \frac{dp}{2\pi} e^{ipz-tH(p)}\\\nonumber
 &=&
\frac{1} {T\,\sqrt{\pi} \Gamma (t)}
\left(\frac{|z|}{2T}\right)^{t-1/2}
K_{t-1/2}(|z|/T).
\label{@KDIS}\end{eqnarray}
 where $t$ is measured in minutes.
For $t=1$ this agrees, of course, with the minute distribution
(\ref{doubleexponential}). \\
The variance of this distribution increases linearly in
time as
\begin{equation}
\sigma^2(t)\equiv\langle z^2\rangle_c(t)=\sigma^2t=2T^2t,~
\label{@TimeDep}\end{equation}
whereas the kurtosis decreases with $1/t$
\begin{equation}
  \kappa  (t)
\equiv \frac{\langle z ^4 \rangle_c(t)}
{\langle z ^2 \rangle_c^2(t)}-3 = \frac{3}{t},~~~~
\label{@TimeDep1}\end{equation}
and goes to zero  for large times
where the
distribution becomes Gaussian.

These quantities are plotted in Figs.~\ref{spvariance} and
\ref{spkappa}. The time dependence of $ \sigma ^2(t)$
in Eq.~(\ref{@TimeDep}) allows us to extract the temperature
 of the
initial Boltzmann distribution
as $T= \sqrt{\sigma ^2(t)/2t}$
 from any later distribution
in which the sharp Boltzmann peak is no longer
visible,
in particular from the asymptotic Gaussian limit.
The
result of this analysis is contained in the plots in
Fig.~\ref{TeMp}. The temperature depends, of course, on the selection of
stocks, but changes only very slowly with the general economic and
political environment.
Near  a crash, however, it increases
significantly.

{\begin{figure}[tbp]
\begin{picture}(1,130)
\put(-130,0){\IncludeEpsImg{194.02mm}{163.09mm}{0.6}{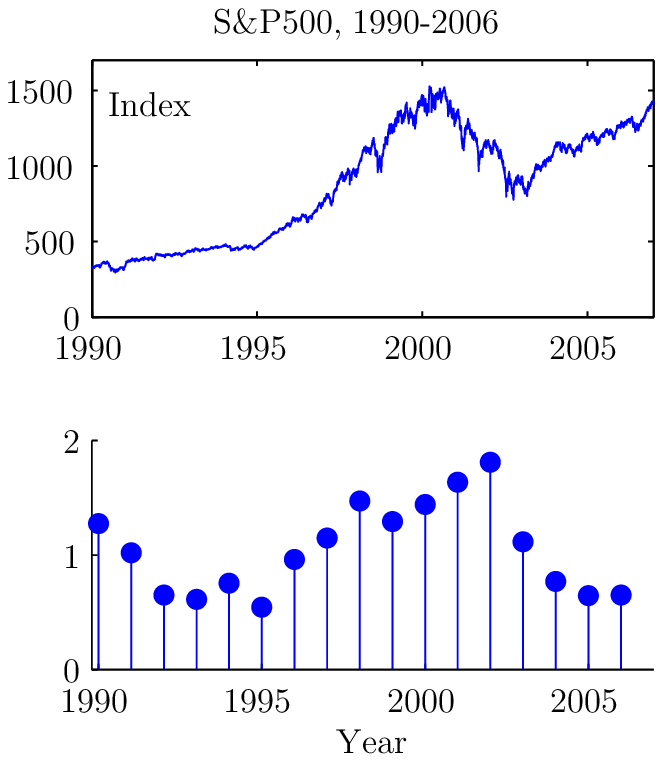}}
\put(-103,50){\tiny{$10^4\,T$}}
\put(10,0){\IncludeEpsImg{194.02mm}{163.09mm}{0.6}{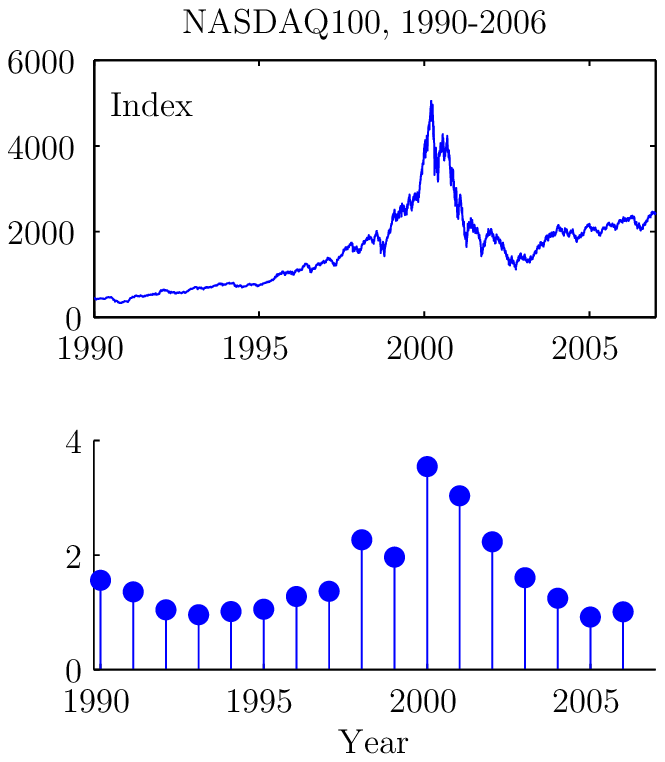}}
\put(37,50){\tiny{$10^4\,T$}}
\end{picture}
~\\[-1em]
\caption{Market temperatures of S\&P 500 and NASDAQ 100 indices from
1990 to 2005. The crash in the year 2000 occurred at the
maximal temperatures
 $T_{\rm S\&P 500}\approx2\times10^{-4}$ and
 $T_{\rm NASDAQ}\approx4\times10^{-4}$.
}
 \label{TeMp}
\end{figure}

It is interesting to observe the
historic development of Dow Jones temperature
over
the last 78
years (1929-2006)
 in Fig.~\ref{dow}.
Although the world went through a lot
of turmoil
and economic development
in the 20th century,
the temperature remained
rather constant except for short heat bursts.
The temperature was highest
in the 1930's, the time of the great
depression. These  temperatures have never been reached
again.
An especially hot burst
occured during
the crash year 1987.

The lesson from this analysis is
that an  increase in market temperature before a crash may
be a useful signal for investors to shorten their positions.

{\begin{figure}[th]
\begin{picture}(1,250)
\put(-150,0){\IncludeEpsImg{194.02mm}{163.09mm}{0.5}{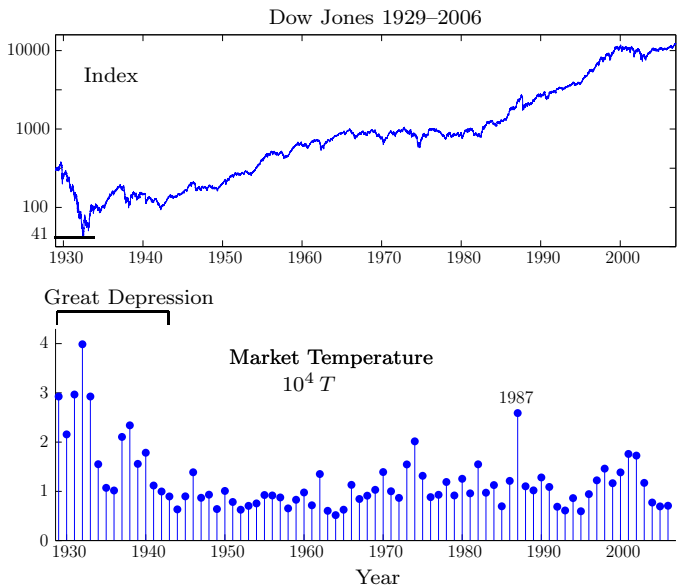}}
\put(-110,108){\line(0,1){5}} \put(-68,108){\line(0,1){5}}
\put(-110,113){\line(1,0){42}} \put(-30,222){\footnotesize{Dow Jones
1929--2006} }\put(-100,200){\footnotesize
Index}\put(-111,141){\line(1,0){15}}\put(-45,93){\footnotesize
Market Temperature }\put(-115,116){\footnotesize Great Depression}
\put(-45,93){\footnotesize Market Temperature }
\put(-25,83){\footnotesize $10^4\,T$} \put(0,10){\footnotesize{
Year}}
\end{picture}
~\\[-1em]
\caption{Dow Jones index over 78 years (1929-2006) and the
yearly
market temperature, which is remarkably uniform, except in
the 1930's, in the beginning of the great depression.
Another heat burst occurred in the crash year 1987
(data from \cite{tick}).
 }
 \label{dow}
\end{figure}


~\\Acknowledgement:\\
The authors are grateful to
 P. Haener and
D. Sornette
for
useful comments.


\begin{thebibliography}{100}


\bibitem{MS1}
R.N. Mantegna and H.E. Stanley, {\em Stochastic Process with
Ultraslow Convergence to a Gaussian:The Truncated L\'evy Flight},
Phys.Rev.Letters {\bf 73}, 2949 (1994); {\em Scaling Behaviour in
the Dynamics of an Economic Index}, Nature {\bf 376}, 46 (1995);
{\em Econophysics: Scaling and Its Breakdown in Finance},
J. Stat. Phys {\bf 89}, 469 (1997);
 {\em An Introduction to Econophysics:
Correlations and Complexity in Finance}, Cambridge University Press,
2000.


\bibitem{BC1}
J.-P. Bouchaud and R. Cont,  {\em Elements for a Theory of Financial
Risk}, Physica A {\bf 263}, 415 (1999); {\em A Langevin Approach to
Stock Market Fluctuations and Crashes}, Eur. Phys. J. B {\bf 6}, 543
(1998)

\bibitem{Pott1}
L. Laloux, M. Potters, R. Cont, J.-P. Aguilar, and J.-P. Bouchaud {\em
Are Financial Crashes Predictable?}, Europhys. Lett. {\bf 45}, 1
(1999)


\bibitem{Pott2}
J.-P. Bouchaud and M. Potters,
 {\em Theory of Financial Risks:\/}
{\em From Statistical Physics to Risk Management\/}, Cambridge
University Press, 2000.


\bibitem{Sornette}

D. Sornette, J. V. Andersen and P. Simonetti,
Int. J. Theor. Appl. Finance {\bf 3}, 523  (2000)
({\tt xxx.lanl.gov/abs/cond-mat/9811292}).

\bibitem{Lux96}
T.~Lux, Appl. Financial Economics {\bf 6}, 463 (1996);
M.~Loretan and P.C.B.~Phillips, J. Empirical Finance {\bf 1}, 211
(1994).

\bibitem{SorMal}

Y. Malevergne, V.F. Pisarenko and D. Sornette,
Quant. Fin. {\bf 5}, 379 (2005)
({\tt arXiv.org/abs/physics/0305089}).

\bibitem{Y1}
A.C. Silva and V.M. Yakovenko, {\em Comparison between the probability
distribution of returns in the Heston model and empirical data for
stock indexes}, physica A {\bf 324}, 303 (2003).


\bibitem{TSALLIS}
C. Tsallis, J. Stat. Phys. {\bf 52}, 479 (1988); E.M.F. Curado and
C. Tsallis, J. Phys. A {\bf 24}, L69 (1991); 3187 (1991); A {\bf
25}, 1019 (1992).

\bibitem{TSALLISN}
C. Tsallis, C. Anteneodo, L. Borland, R. Osorio, {\em Nonextensive
Statistical Mechanics and Economics\/}, Physica A {\bf 324}, 89
(2003) (cond-mat/030130);
 L. Borland, Phys. Rev. Lett. {\bf 89}, 098701 (2002), and references therein.

\bibitem{PJ}
W.Paul and J.Baschnagel,
 {\em Stochastic Processes:\/}
{\em From Physics to Finance\/}, Springer, 2000.

\bibitem{Voigt}
J. Voit, {\em The Statistical Mechanics of Financial Markets\/},
Springer, Berlin, 2001.

\bibitem{PI}
 H. Kleinert,
     {\em Path Integrals in Quantum Mechanics, Statistics,
 Polymer Physics, and Financial Markets\/},
     World Scientific, Singapore 2004,
     Third extended edition, pp. 1--1450
({\tt www.physik.fu-berlin.de/{}\~{}{}kleinert/b5}).




\bibitem{Go+98} P.~Gopikrishnan, M.~Meyer, L.A.N.~Amaral, and H.E.~Stanley,
Eur. Phys. J. B {\bf 3}, 139 (1998).





\bibitem{EXPOD}
In the literature the exponential distribution
is sometimes referred to as
 Laplace distribution. On also finds the name
double-exponential distribution
which emphasizes the fact that
there are
two in general different branches, one for positive and one for negative returns,
which we equated in this paper,  for simplicity.



\bibitem{Y2}
A.C. Silva, R.E. Prange, and V.M. Yakovenko, {\em Exponential distribution
of financial returns at mesoscopic time lags:a new stylized fact},
physica A {\bf 344}, 227 (2004).


\bibitem{Y4}
A.C. Silva and V.M. Yakovenko, {\em
Stochastic volatility of
ÿ ÿ financial markets as the fluctuating rate of trading: an empirical
ÿ ÿ study\/},
(physics/0608299).

\bibitem{Yahoo}
Our data
are taken from
 {\tt www.tickdata.com}.

\bibitem{Aoki}
M. Aoki,
{\em New Approaches to Macroeconomic Modeling\/},
Cambridge University Press, Cambridge, 1996.

\bibitem{Y3}

A. Dragulescu and V.M. Yakovenko,
Eur. Phys. J. B {\bf 17}, 723 (2000).










\bibitem{tick}
  Yahoo Finance {\tt finance.yahoo.com}. To download
   data, enter in the symbol box: {\tt \^{}DJI}, and then click on the
   link: {\em Download Spreadsheet\/}.

\end{thebibliography}
\end{document}